\pdfoutput=1
\documentclass[twocolumn,
               showpacs,
               preprintnumbers,
               footinbib,
               prl,
               superscriptaddress,
               10pt,
               notitlepage,
               aps]{revtex4-2}
               
\usepackage{graphicx,amssymb,amsmath,amsthm,amsfonts,epsfig,mathtools}

\usepackage[utf8]{inputenc}
\usepackage{graphicx}
\usepackage{float}
\usepackage{dcolumn}
\usepackage{bm}
\usepackage{color}
\usepackage{soul}
\usepackage[dvipsnames]{xcolor}
\usepackage{hyperref}
\hypersetup{colorlinks=true, citecolor=MidnightBlue,linkcolor=CornflowerBlue, urlcolor=CornflowerBlue, linktocpage=true}
\usepackage{xfrac}
\usepackage{siunitx}

\newcommand{\beqn}{\begin{eqnarray}}
\newcommand{\eeqn}{\end{eqnarray}}
\newcommand{\beq}{\begin{equation}}
\newcommand{\eeq}{\end{equation}}
\newcommand{\thetitle}{Nature of phase transitions and metastability in scalar-tensor theories}

\def\d{\mathrm{d}}

\def\mphi{m_{\phi}}

\begin{document}

\title{\thetitle}

\begin{abstract}
Compact stars above a critical stellar mass develop large scalar fields in some scalar-tensor theories. This scenario, called spontaneous scalarization, has been an intense topic of study since it passes weak-field gravity tests naturally while providing clear observables in the strong-field regime. The underlying mechanism for the onset of scalarization is often depicted as a second-order phase transition. Here, we show that a first-order phase transition is in fact the most common mechanism. This means metastability and transitions between locally stable compact object configurations are much more likely than previously believed, opening vast new avenues for observational prospects.
\end{abstract}

\author{K{\i}van\c{c} \.I. \"Unl\"ut\"urk}
\email{kivanc.unluturk@tau.edu.tr}
\affiliation{Department of Electrical and Electronics Engineering, Turkish-German University,\\
34820 Beykoz, \.{I}stanbul, T\"{u}rkiye}
\affiliation{Department of Physics, Ko\c{c} University, Rumelifeneri Yolu,\\
34450 Sar{\i}yer, \.{I}stanbul, T\"{u}rkiye}

\author{Semih Tuna}
\email{semih.tuna@columbia.edu}
\affiliation{Department of Physics and Columbia Astrophysics Laboratory, Columbia University, New York, NY 10027, USA}

\author{O\u{g}uzhan K. Yamak}
\email{oyamak21@ku.edu.tr}
\affiliation{Department of Physics, Ko\c{c} University, Rumelifeneri Yolu,\\
34450 Sar{\i}yer, \.{I}stanbul, T\"{u}rkiye}

\author{Fethi M. Ramazano\u{g}lu}
\email{framazanoglu@ku.edu.tr}
\affiliation{Department of Physics, Ko\c{c} University, Rumelifeneri Yolu,\\
34450 Sar{\i}yer, \.{I}stanbul, T\"{u}rkiye}

\date{\today}
\maketitle

Advances in gravitational and electromagnetic observations now enable us to test general relativity (GR) in unprecedented ways~\cite{LIGOScientific:2016lio,LIGOScientific:2018dkp,KAGRA:2021duu,KAGRA:2021vkt, LISA:2022kgy,Barack:2018yly,Gendreau2017}. The particular scenario of \emph{spontaneous scalarization} in scalar-tensor theories (STTs) has been a prime target of study for deviations from GR, as it readily satisfies the strict observational limits in weak fields while also promising achievable observational targets in strongly gravitating systems~\cite{Damour:1993hw,Doneva:2022ewd}. Hence, the resulting literature has been vast and impactful~\cite{Damour:1996ke,Sperhake:2017itk,Doneva:2017bvd, Silva:2017uqg,Mendes:2018qwo,Herdeiro:2018wub,Herdeiro:2020wei,Berti:2020kgk,Kuan:2022oxs}. Here, we report that a first-order phase transition is a considerably more likely mechanism for the onset of scalarization, unlike the commonly invoked second-order phase transition picture. This means distinct phenomena such as metastable neutron star (NS) configurations and discontinuous jumps between them are more likely than previously thought. This, in turn, opens new avenues for both theory and observation.

In spontaneous scalarization, gravity is governed by a fundamental scalar field in addition to the metric tensor, and the scalar field vacuum can become unstable in the highly curved regions of spacetime~\cite{Doneva:2022ewd}. While ordinary stars and low-mass NSs can behave as GR dictates, a scalar cloud forms around the NS if its mass exceeds a critical value. This is commonly explained in analogy to spontaneous magnetization, and we have a \emph{second-order phase transition} at the critical mass~\cite{Damour:1996ke}. The scalar field strength continuously rises from zero with increasing stellar mass, and can have large enough values to cause nonpertubative deviations from GR that are relatively easy to observe.

We show that the above picture only holds for a very limited part of the STT parameter space, and the onset of scalarization occurs as a discontinuous jump in the stellar structure in most cases. This means scalarized and unscalarized solutions are both possible for some stellar masses, one of the solutions being the globally stable one while the other is metastable. All these findings are successfully explained as a first-order phase transition, which also elucidates many previously opaque numerical results.

Metastability in scalarization has been investigated in some particular contexts before~\cite{Kuan:2022oxs,Doneva:2021tvn, Doneva:2022yqu, Doneva:2023kkz, Pombo:2023lxg, Staykov:2024jbq}, but our results imply that first-order phase transitions are the rule rather than the exception. This means limited existing studies have much wider applicability than previously considered, providing novel ways to test gravity.

\noindent {\bf \em  The theory:}
Consider the STT action
\begin{align}
    S = &\frac{1}{16\pi} \int \d^4 x \sqrt{-g} \left(R - 2g^{\mu\nu} \nabla_\mu\phi \nabla_\nu\phi - 2m_{\phi}^2 \phi^2 \right) \nonumber\\
    & + S_{\text{m}}[f_{\text{m}}, \tilde{g}_{\mu\nu}= A^2(\phi)g_{\mu\nu}]. 
    \label{eq: action}
\end{align}
$S_{\text{m}}$ is the matter action and $f_{\text{m}}$ denotes the matter fields which couple to the conformally scaled \emph{Jordan-frame metric}. We use $A(\phi)=e^{\beta \phi^2/2}$ as in the original discovery paper of \textcite{Damour:1993hw}, and add the scalar mass term $\mphi$~\cite{Ramazanoglu:2016kul}. Hence, our STTs live on a $(\beta,\mphi)$ parameter space. Note that $\mphi$ has the important consequence of suppressing the scalar dipole radiation~\cite{Alsing:2011er,Ramazanoglu:2016kul,Doneva:2022ewd}, which means our model here is not affected by various observations that severely constrain the original massless scalarization idea~\cite{2013Sci...340..448A,Kramer:2021jcw,Zhao:2022vig}.

The scalar field obeys
\begin{equation}\label{eq: scalar field eqn for given A}
    \Box \phi = \left(-4\pi\beta A^4 \tilde{T} + m_\phi^2 \right) \phi,
\end{equation}
$\tilde{T}$ being the trace of the stress-energy tensor in the Jordan frame. The term in the parentheses becomes negative at high matter densities for $\beta<0$, which leads to a mode with purely imaginary oscillation frequency, a tachyon. Scalar fields grow exponentially around massive NSs due to this instability, which is eventually suppressed by nonlinear effects, leading to a star inside a stable scalar cloud, called a \emph{scalarized star}. Note that an unscalarized star with $\phi=0$, a GR solution, is always an equilibrium solution for our choice of $A(\phi)$, but it is not stable if the tachyon is present.

\noindent {\bf \em   Metastability in spontaneous scalarization:}
%
\begin{figure}
    \centering
    \includegraphics[width=\columnwidth]{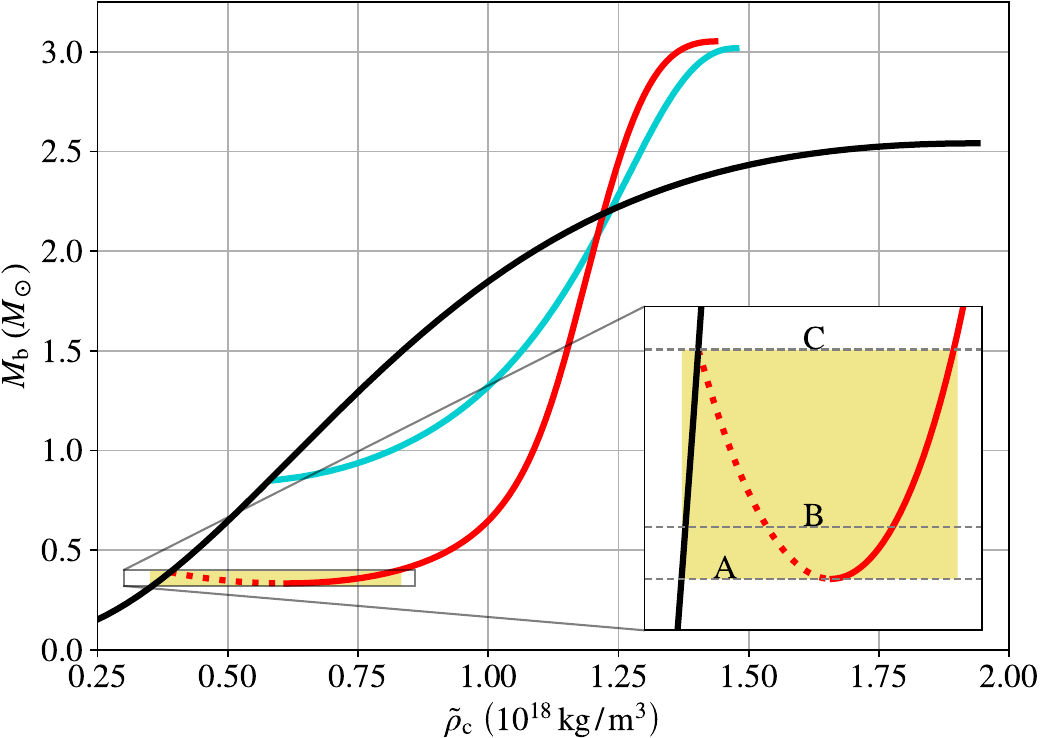}
    \caption{Baryon mass vs the central density $\tilde{\rho}_\text{c} = \tilde{\rho}(r{=}0)$ diagrams under general relativity (black) and STTs with spontaneous scalarization (red, turquoise). Turquoise: $\beta=-10$, $m_\phi=1\times10^{-11}\,\text{eV}$. Red: $\beta=-40$, $m_\phi = 4\times 10^{-11}\,\text{eV}$.}
    \label{fig: m-rho_c curves with inset}
\end{figure}
We used the numerical techniques detailed in \textcite{Tuna:2022qqr} to obtain the static, spherically symmetric NS solutions for the action~\eqref{eq: action}, also see the ``End Matter.'' Both scalarized and unscalarized solutions are plotted in Fig.~\ref{fig: m-rho_c curves with inset} as one-parameter families in terms of the central density of the star $\tilde{\rho}_\text{c}$. We use the HB equation of state (EOS) of \textcite{Read:2009yp} in all cases, but results are qualitatively similar for other choices.

Our main objective is understanding the physical relevance of the solutions when there exists more than one configuration for a given number of baryons $N_\text{b}$ in the star, or equivalently the total baryon mass $M_\text{b} = m_\text{b}N_\text{b}$, where $m_\text{b}$ is the rest mass of a single baryon. In other words, if we are given a fixed number of baryons and there are more than one theoretical equilibrium configuration they can arrange themselves in, which one(s) are astrophysically relevant?

Let us start with the ``standard'' scalarization picture. The unscalarized solutions, which are also solutions in GR, are on the black curve, and the scalarized ones are on the turquoise one in Fig.~\ref{fig: m-rho_c curves with inset}. Scalarized solutions branch off at a critical baryon mass $M_\text{crit}$, below which there is only one solution for a given baryon mass; an unscalarized one. This solution is known to be stable, hence, can be observed in principle. For $M_\text{b} > M_\text{crit}$ though, there are two possible solutions, one of them being scalarized~\footnote{There is a $\phi \to -\phi$ symmetry for our choice of $A(\phi)$, which means each scalarized solution in the figure represents two configurations with opposite scalar field parity. We will be talking about points on the scalarized branches of the $M_\text{b}(\tilde{\rho}_\text{c})$ curve as if they are single solutions to avoid confusion, since they look as such on the $M_\text{b}(\tilde{\rho}_\text{c})$ curves.}. However, we have seen that the unscalarized solution is unstable to the growth of a scalar mode, hence it would not be observed. Thus, any existing star with these higher masses would be a scalarized one. We call this picture \emph{second-order scalarization.} The naming will become clear in our phase transition discussion.

Our main interest is the \emph{first-order scalarization} of the red curve in Fig.~\ref{fig: m-rho_c curves with inset}, which corresponds to different theory parameters $(\beta, \mphi)$. The scalarized section of the $M_\text{b}(\tilde{\rho}_\text{c})$ curve slopes downwards at the branch-off point $M_\text{crit}$ in this case, which changes the above discussion substantially. Defining the lowest mass on the scalarized branch as $M_\text{bottom}$, there is a single scalarized and stable equilibrium solution for $M_\text{b} < M_\text{bottom}$. There are two solutions for $M_\text{b} > M_\text{crit}$, where the unscalarized solution (black) is unstable due to the tachyonic growth, and the scalarized one (red) is stable. This is similar to second-order scalarization so far.

The novelty is in the highlighted region $M_\text{bottom} < M_\text{b} < M_\text{crit}$ in Fig.~\ref{fig: m-rho_c curves with inset}, where there are three equilibrium solutions for a given $M_\text{b}$. Firstly, the middle solution on the dotted curve, is unstable, hence physically irrelevant. A one parameter family of equilibrium NS solutions, such as our $M_\text{b}(\tilde{\rho}_\text{c})$, change from being stable to unstable at a turning point $\d M_\text{b} / \d \tilde{\rho}_\text{c} = 0$, and the  $\d M_\text{b} / \d \tilde{\rho}_\text{c} < 0$ part is known to be the unstable one~\cite{Sorkin:1981jc,Sorkin:1982ut,Shapiro:1983du,Harada:1998ge,Mendes:2018qwo}. This instability is not necessarily due to the tachyonic scalar mode, but can also be a result of the modes of collective motion of the stellar matter.

The remaining two solutions, the scalarized star with $\d M_\text{b} / \d \tilde{\rho}_\text{c} > 0$ (solid red) and the unscalarized star (black), are at least locally stable. The former is stable by being on the other side of a turning point, and the latter because there is no hydrodynamical instability on this part of the GR curve, and the tachyonic instability of the scalar is not present before $M_\text{crit}$. Thus, we have two physically relevant solutions sharing the same $M_\text{b}$.

When there are multiple equilibrium solutions that are robust against small perturbations, as is our case, we have the phenomenon of \emph{metastability}. We typically call the solution with the globally lowest energy \emph{the} stable one (or the global minimum), whereas the others are metastable. A metastable state would energetically prefer to transition to the global minimum, but this can only occur for large enough perturbations which may or may not exist in a star's environment. Hence, both solutions are possible in reality, which makes novel astrophysical scenarios possible as we will discuss later.

\begin{figure}
    \centering
    \includegraphics[width=\columnwidth]{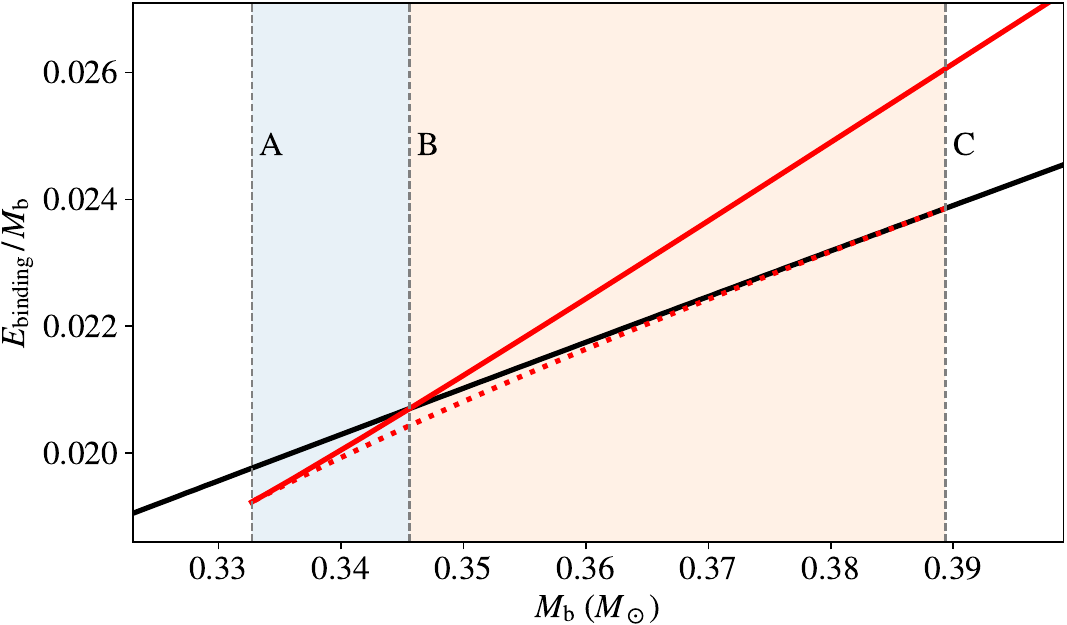}
    \caption{Fractional binding energy $(M_\text{b} - M_\text{ADM})/M_\text{b}$ in the highlighted region of Fig.~\ref{fig: m-rho_c curves with inset}.}
    \label{fig: binding energy atypical}
\end{figure}
The total energy of a NS is its ADM mass $M_\text{ADM}$. Hence, we can determine metastability by checking the total binding energy $E_{\rm binding} = M_\text{b} - M_\text{ADM}$ plotted in Fig.~\ref{fig: binding energy atypical}, which shows that metastability is not exclusive to one branch of solutions. For lower $M_\text{b}$ (to the left of the vertical line B) the unscalarized (black) configuration is energetically favored, and the opposite is true for higher $M_\text{b}$. We also see that the unstable scalarized solutions (dotted red) always have higher energy than the locally stable ones.

\begin{figure}
    \centering
    \includegraphics[width=\columnwidth]{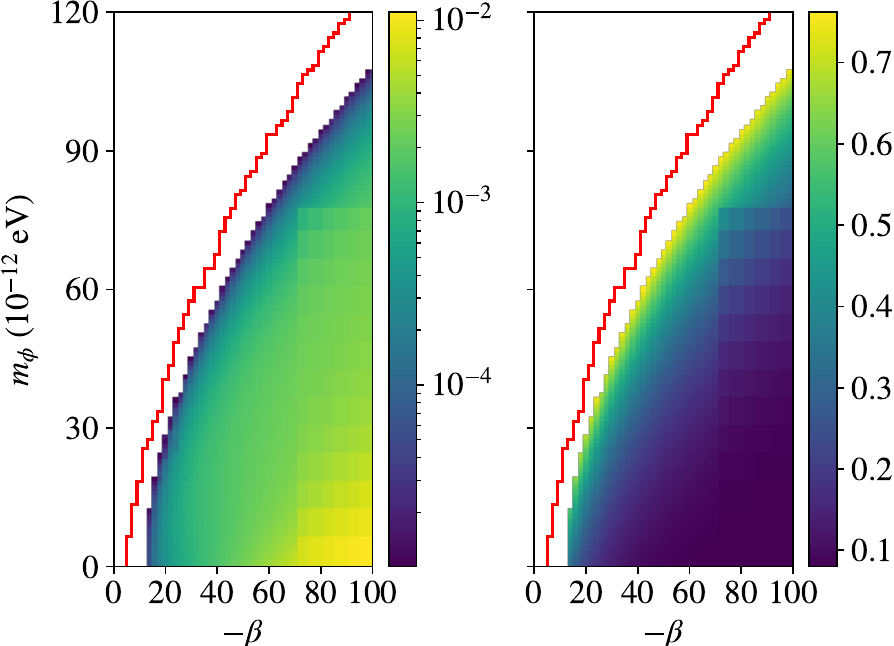}
    \caption{Dependence of scalarization characteristics on the theory parameters $(\beta,\mphi)$ for first order scalarization. There is no scalarization at all above the red line~\cite{Tuna:2022qqr}. Second-order scalarization occurs only in the white regions below the red line. Thus, first-order scalarization is by far the more likely outcome if scalarization occurs. Left: $\Delta M$, the maximum ADM mass difference between two stars of equal baryon mass. Right: ADM mass at the scalarization branch-off point. Both are in units of $M_\odot$. $\Delta M<10^{-5}M_\odot$ is indistinguishable from numerical noise, hence excluded from the plot. We use lower resolution away from the boundary between first- and second-order scalarization, where fine detail is not critical. See ``End Matter'' for details.}
    \label{fig: energy differences HB EOS}
\end{figure}

We closely investigated only two points on the $(\beta,\mphi)$ parameter space of our STT so far, but what is the prevalence of first- and second-order scalarization in general? Fig.~\ref{fig: energy differences HB EOS} (left) shows the maximum ADM mass difference $\Delta M$ between two stars with the same baryon mass, which is a crude approximation for the energy that would be released in a transition between two states where there is no matter loss. Since it means there are metastable states, any point with a nonzero value means scalarization is first order if the STT with parameters $(\beta,\mphi)$ governs gravity. One can see that first-order scalarization is the dominant type even for the original scalarization model for which $\mphi=0$~\cite{Damour:1993hw}. This clearly demonstrates our claim about the ubiquity of first-order scalarization.

Fig.~\ref{fig: energy differences HB EOS} (right) shows the ADM mass at the branch-off point of scalarization, $M_\text{ADM,branch}$ only for the parameter values featuring first-order scalarization. This is relevant since it gives us a rough measure of the NS (ADM) masses where transitions from a metastable to a stable configuration can occur. Metastable solutions mostly have $M_\text{ADM,branch} \lesssim 0.8 M_\odot$, which is low but astrophysically relevant~\cite{Doroshenko:2022nwp}. We will further comment on this in our discussion.

In terms of detectability, relatively higher NS masses and higher $\Delta M$ are both desirable. This means that the prominent regions in Fig.~\ref{fig: energy differences HB EOS} are not necessarily interesting on their own, but rather the middle region where $\Delta M$ and $M_\text{ADM,branch}$ are both relatively high likely provides the best observational prospects. Consider these results in the context that they are only for a specific STT. $\Delta M$ and $M_\text{ADM,branch}$ would vary for other, more general, STTs, providing opportunities for testing deviations from GR.

\noindent {\bf \em   Phenomenological descriptions of scalarization:}
The onset of scalarization has been viewed as a phase transition since the early days~\cite{Damour:1996ke}, and phenomenological explanations in terms of Landau theory have been occasionally invoked in the literature~\cite{Harada:1998ge,Khalil:2019wyy}. Nevertheless, this has been in the context of second-order phase transitions, which only holds for second-order scalarization~\footnote{\textcite{Muniz:2025egq} is an exception, which was posted after the submission of our manuscript.
The end, rather than the beginning, of scalarization was called a first-order phase transition in \textcite{Kuan:2022oxs}, though this study was not specifically concerned with a phenomenological analysis.}. Here, we show that all unusual features of first-order scalarization in the $M_\text{bottom} < M_\text{b} < M_\text{crit}$ region can be explained by a first-order phase transition.

In Landau theory~\cite{Plischke2006, Goldenfeld2018}, one considers how the total free energy of a system changes with a quantity called the \emph{order parameter}, based on the symmetries of the system. In our case, we will concentrate on the simpler $\mphi=0$ case where this translates to
\begin{equation}
\label{eq: energy landau ansatz}
    M_\text{ADM} = M_0(M_\text{b}) + a(M_\text{b}) Q^2 + \frac{1}{2} b(M_\text{b}) Q^4 +  \frac{1}{3} c(M_\text{b}) Q^6 .
\end{equation}
The \emph{scalar charge} of the star, defined as $\phi(r \to \infty) = \phi_\infty + Q/r + \cdots$ has been used as the order parameter $Q$~\cite{Damour:1993hw, Damour:1996ke}\footnote{We are only interested in solutions with $\phi_\infty=0$, it therefore plays no role in our phase transition picture. However, in general, $\phi_\infty$ has an effect on scalarization analogous to the effect of an external magnetic field on magnetization.}. In rough terms, we take a given amount of baryons $M_\text{b}$, and determine how the total energy of the solution changes as we dress it with scalar fields whose strength is represented by $Q$. The scalar charge is not well defined for massive scalars, but some other measure of the strength of scalarization such as the scalar field value at the center can be used in a similar vein. Only even terms are present in Eq.~\eqref{eq: energy landau ansatz} due to the $\phi \to -\phi$ symmetry. $c(M_\text{b})>0$ for overall stability of the system.

Second-order scalarization occurs when $b(M_\text{b}) > 0$. We overlook $c(M_\text{b})$ which plays no direct role in this case. The two variables $M_\text{b}$ and $Q$ are actually not independent for the equilibrium solutions, i.e., the static NS solutions we studied above. For a given $M_\text{b}$, the physically realized $Q$ will be the one that minimizes the total energy $M_\text{ADM}$. For $a>0$, this is simply $Q=0$, the unscalarized solution. For $a<0$, however, $Q=0$ becomes a local maximum, an unstable equilibrium, and we get two minima at $Q=\pm(-a/b)^{1/2}$. Thus, as $a$ changes sign, the $Q\leftrightarrow-Q$ symmetry gets spontaneously broken, and a phase transition to nonzero values of $Q$ occurs \emph{continuously:} physically relevant equilibria with arbitrarily small $Q$ are possible. Recall that both signs of $Q$ correspond to the same $M_\text{b}(\tilde{\rho}_\text{c})$ curve in Fig.~\ref{fig: m-rho_c curves with inset}.

This is a textbook example of a \emph{second-order phase transition,} hence our nomenclature. In particular, $a(M_\text{b})$ changes sign at a critical baryon mass where stable scalarized solutions appear, which is the $M_\text{crit}$ from the previous section. This phenomenological explanation goes back to the earliest works on the topic~\cite{Damour:1996ke,Harada:1998ge}.

\begin{figure}
    \centering
    \includegraphics[width=\columnwidth]{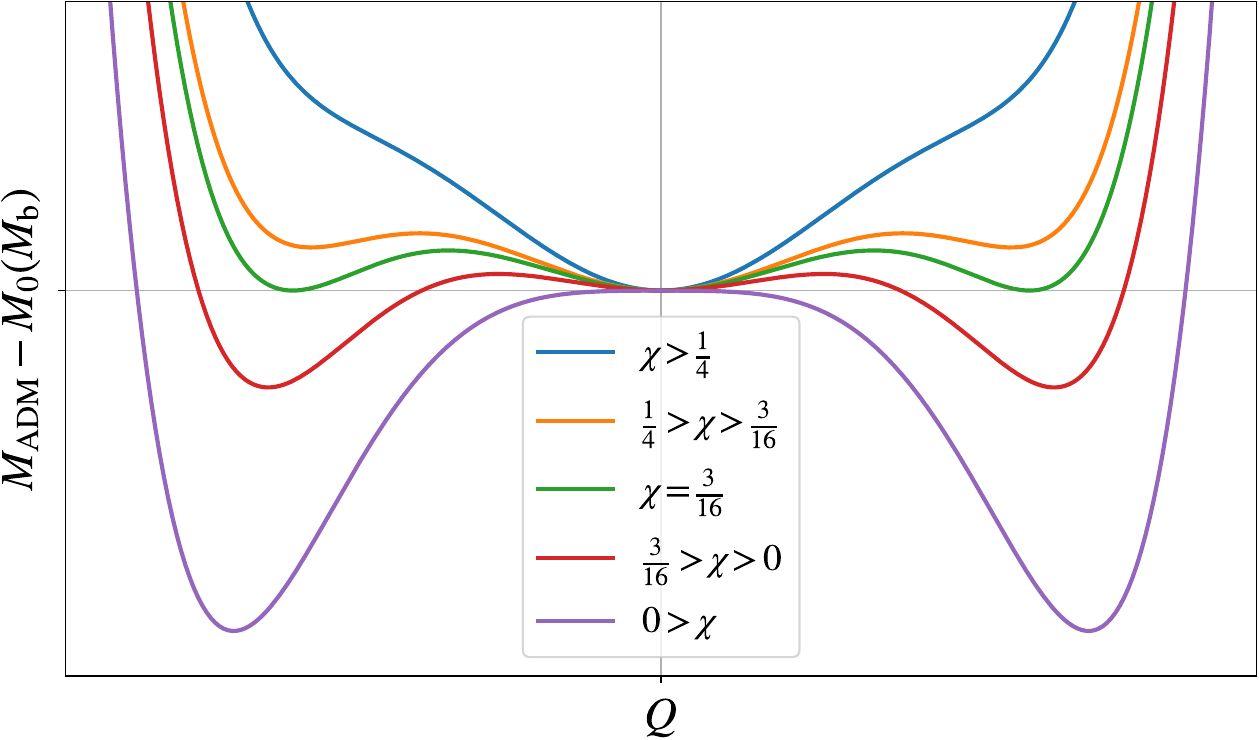}
    \caption{The Landau ansatz in Eq.~\eqref{eq: energy landau ansatz} in the case $b<0$, plotted for different values of the parameter $\chi \equiv ac/b^2$. This is the case of a first-order phase transition.}
    \label{fig:first_order}
\end{figure}
First-order scalarization occurs when $b(M_\text{b}){<}0$, for which $c(M_\text{b}){>}0$ is also relevant. The behavior of $M_\text{ADM}(Q)$ changes drastically depending on the parameter $\chi \equiv ac/b^2$ that decreases with increasing $M_\text{b}$, as seen in Fig.~\ref{fig:first_order}. There is a single stable equilibrium at $Q=0$ when $\chi$ is positive and high, which becomes unstable for negative $\chi$ for which two symmetric stable equilibria exist. In between these phases, we have three local minima corresponding to stable or metastable states, as well as two symmetric local maxima that correspond to unstable equilibria. Quantitative details can be found in the ``End Matter.''

The phase transition picture in Fig.~\ref{fig:first_order} elucidates many points in our numerical results which previously had no clear reason. For one, the unstable scalarized stars always have the lowest binding energy in Fig.~\ref{fig: binding energy atypical}, even though there is no necessity for unstable equilibria to be energetically disfavored. This becomes a trivial point in Fig.~\ref{fig:first_order} since a local maximum between two local minima has to have the highest energy. The list continues for other features like (i) the stable scalarized stars being metastable at lower $M_\text{b}$ and becoming the global minimum only with increasing $M_\text{b}$, (ii) the branch continuously connected to the GR solutions being unstable and (iii) the unstable scalarized stars always having lower scalar charge compared to the scalarized ones: $Q_- < Q_+$ (not plotted here).

The main factor controlling the order of the phase transition is the sign of the coefficient $b(M_\text{b})$ in the Landau ansatz~\eqref{eq: energy landau ansatz}. For various theory parameters, we extracted the value of $b$ at the point where the scalarized NSs branch off from the GR solutions, $b_0=b(M_\text{crit})$. See ``End Matter'' for details.

\begin{figure}
    \centering
    \includegraphics[width=\columnwidth]{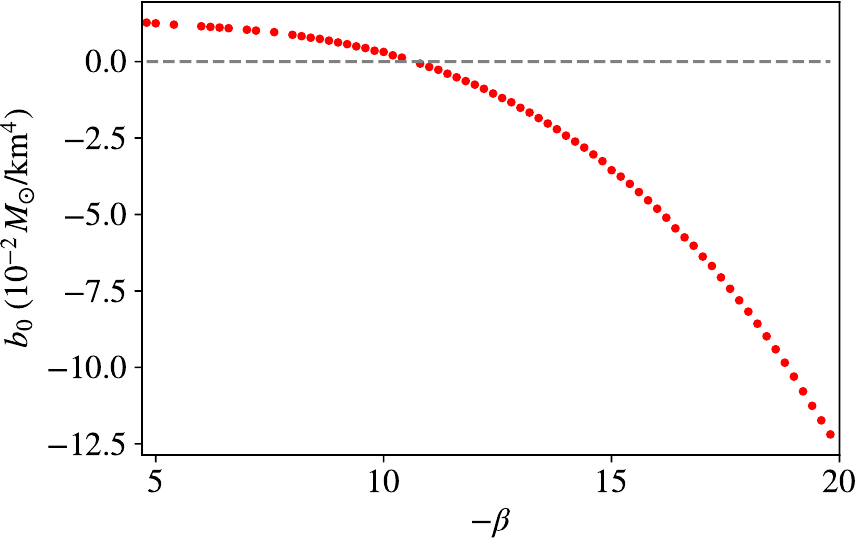}
    \caption{$b_0=b(M_\text{crit})$ as a function of $\beta$ for $\mphi=0$. The change from first- to second-order scalarization is near $\beta=-10.6$, that is, $b_0(\beta=-10.6)=0$.}
    \label{fig:b_vs_beta}
\end{figure}

Fig.~\ref{fig:b_vs_beta} clearly shows how $b_0$ becomes more negative as $\beta$ becomes more negative for $\mphi=0$, the case being similar for all scalar masses. This is exactly in line with our observations in Figs.~\ref{fig: energy differences HB EOS} and~\ref{fig:first_order}.
%

\noindent {\bf \em  Discussion:}
We have demonstrated that first-order scalarization is the norm for our choice of STT action~\eqref{eq: action}, but we also expect this to be a common occurrence. Scalarization is now known to be a generic phenomenon for various possible couplings between a scalar field and the metric~\cite{Doneva:2017bvd,Silva:2017uqg,Andreou:2019ikc,Doneva:2022ewd}. There are negative cases, as our preliminary analysis indicates that the coupling choice $A(\phi)=\cosh\left(\sqrt{3}\beta \phi \right)^{1/3\beta}$~\cite{Mendes:2018qwo,Demirboga:2023ktt} features first-order scalarization less prominently. On the other hand, some other STTs feature first-order transitions even more prominently than Eq.~\eqref{eq: action}. For example, scalarization of both black holes and NSs in scalar-Gauss-Bonnet theories are known to preferentially have first-order scalarization when couplings are explicitly enhanced in higher orders, e.g. $A(\phi)=e^{\beta \phi^2/2+\gamma \phi^4/4}$~\cite{Doneva:2021tvn, Doneva:2023kkz, Doneva:2022yqu, Pombo:2023lxg, Staykov:2024jbq}.

Our phase transition picture already has a simple and powerful explanation for the above results. Changing the $\phi^4$ term in the expansion of $A(\phi)$ directly modifies the higher order expansion coefficients in our Landau ansatz~\eqref{eq: energy landau ansatz}, specifically $b$. Recalling that the main difference between a first- and second-order transition is the sign of $b$, we can see how these higher order terms can determine the order of the phase transition. However, we emphasize that explicitly modifying the coupling or some form of \emph{fine tuning} is not necessary for first-order scalarization by any means, as is clear from our choice of $A(\phi)$ here. Overall, we aim to study first-order scalarization for the most general STTs in order to devise more comprehensive tests for these theories. 

In terms of observational prospects, the specific examples we gave are most relevant for lower mass NSs since we concentrated on the onset of scalarization. There are recently proposed theoretical scenarios for these such as formation of NSs down to masses of $0.1M_\odot$ in accretion disks~\cite{Metzger:2024ujc}, and tidal disruption events in black hole--NS mergers where part of the NS mass can be ripped out~\cite{Stephens:2011as,East:2011xa}. Such a change in the stellar mass can radically affect how large a scalar field can be supported by a star in general, and this becomes especially relevant for first-order scalarization where the discontinuous nature of the scalarization/descalarization process can provide distinct signals. There is already some work about first-order phase transitions near the maximum allowed NS masses, for which possible astrophysical signals have been studied~\cite{Kuan:2022oxs}. For instance, a sudden transition between scalarized and non-scalarized configurations results in the emission of gravitational waves which carry two additional modes beyond those in GR.  Moreover, these can also persist for longer periods. Together with related electromagnetic and neutrino signals from such a violent event, these gravitational waves offer distinct observational indicators for a first-order transition. Due to its discontinuous nature, first-order scalarization might also feature novel phenomenology in stellar collapse events, which have only been studied for the continuous second-order scalarization so far~\cite{Sperhake:2017itk}. 

We reiterate that action~\eqref{eq: action} is the original and most popular scalarization theory in terms of the scalar coupling function $A(\phi)$, but it is still one specific choice. We expect other models of scalarization to have diverse characteristics, where different NS mass ranges are possible or even dominant for first-order scalarization. Studying these more general theories and how to test them via discontinuous phase transitions are major future research directions. Also note that the mass term $\mphi$ makes almost all of our parameter space, the region with $\mphi \gtrsim 10^{-15}$eV, unconstrained with current observations~\cite{Ramazanoglu:2016kul,Doneva:2022ewd}. We included the mostly-constrained $\mphi=0$ case in our analysis for the sake of completeness and also because it is still informative since the $\mphi \to 0$ limit is continuous in terms of the NS structure.

The phase transition order depends on the sign of $b$. Let us remark that the parameters $a,b,c$, as well as the function $M_0$ and the critical mass $M_\text{crit}$ depend on the parameters of the theory, $(\beta, m_\phi)$, and also the EOS. For simplicity let us focus on the $\beta$ dependence, i.e. $b = b(M_\text{b},\beta)$. The fact that the order of the phase transition changes at some point as $\beta$ becomes more and more negative means that $b$ changes sign as $\beta$ drops below a critical value $\beta_\text{crit}<0$, see Fig.~\ref{fig:b_vs_beta}. When $\beta=\beta_\text{crit}$, we have a point where $a$ and $b$ vanish simultaneously, i.e., the point $(\beta=\beta_\text{crit}, M_\text{b} = M_\text{crit}(\beta_\text{crit}))$. Such a point is called a \emph{tricritical point} \cite{Plischke2006, Goldenfeld2018}, and it separates a line of second-order phase transitions on the $(\beta, M_\text{b})$ plane from a line of first-order ones. Spontaneous scalarization at $\beta_\text{crit}$ features special characteristics whose study is another possible future research topic. Similarly, other aspects of the voluminous phase transition literature can be a source of other surprising results for gravitational phase transitions.

Lastly, we note that even though a phenomenological fit with very few parameters is possible, there is no understanding for the behavior of $b_0$ in Fig.~\ref{fig:b_vs_beta} from first principles to the best of our knowledge. Recall that the scalar field profiles depend on the nonlinear terms that suppress the tachyonic instability, hence there is no simple method to relate the energy of the spacetime and the related Landau ansatz parameters to the STT parameters. Qualitatively, first-order scalarization becomes dominant when scalarization itself is stronger in terms of how much deviation we have from GR, which happens at higher $|\beta|$~\cite{Ramazanoglu:2016kul,Doneva:2022ewd}. Whether this can lead us to the underlying reason for the change of sign in $b_0$ remains to be seen.

This study was supported by the Scientific and Technological Research Council of Turkey (T\"UB\.ITAK) Grant Number 122F097. Authors thank T\"UB\.ITAK for its support, and Daniela Doneva and Stoytcho Yazadjiev for helpful discussions. Computations were performed on the KUACC cluster of Ko\c{c} University.

%

\hfill\\

\onecolumngrid
\begin{center}
	\textbf{\large End Matter}
\end{center}
\twocolumngrid
\setcounter{equation}{0}
\setcounter{figure}{0}
\setcounter{table}{0}

\renewcommand{\thefigure}{E\arabic{figure}}
\setcounter{figure}{0}

\renewcommand\theequation{E\arabic{equation}}

\noindent {\bf \em Computing the equilibrium NS solutions:}
A static, spherically-symmetric metric can be put in the form
\begin{equation}
\label{eq: spherically symmetric metric}
    g_{\mu\nu} \d x^\mu \d x^\nu = - e^{\nu(r)} \d t^2 + \frac{\d r^2}{1-2\mu(r)/r} + r^2 \d\Omega^2,
\end{equation}
NS structures in this case can be obtained by solving the modified Tolman–Oppenheimer–Volkoff (TOV) equations. Details can be found in \textcite{Tuna:2022qqr}.

One needs to know the nuclear matter equation of state (EOS) to solve the TOV equations. A simple choice is a polytrope that relates the pressure of nuclear matter to the \emph{rest mass density} of baryons $\tilde{\rho}_\text{r}$ (in the Jordan frame, i.e., in the frame of the metric $\tilde{g}_{\mu\nu}$)
\begin{subequations}
\begin{align}
    \tilde{\rho} = \tilde{\rho}_\text{r} + \frac{C}{\Gamma-1} \tilde{\rho}_\text{r}^\Gamma,\ \ \ \
    \tilde{p} = C \tilde{\rho}_\text{r}^\Gamma ,
\end{align}
\end{subequations}
where $C$ and $\Gamma$ are constants. Note that the two equations are related by the first law of thermodynamics. This relationship is sometimes expressed in terms of the number density of baryons $\tilde{n}$, which is linearly related to the baryon rest mass density as $\tilde{\rho}_\text{r} = m_{\rm b} \tilde{n}$. In this study, we employ the slightly more complicated \emph{piecewise polytropic} EOS introduced in \textcite{Read:2009yp}.

Once the TOV equations are solved, the total energy of the spacetime, the Arnowitt-Deser-Misner (ADM) mass $M_\text{ADM}$, is given by 
\begin{align}
    M_\text{ADM} = \mu(r \to \infty) .
\end{align}
\emph{Baryon mass}, which is the integral of $\tilde{\rho}_\text{r}$ over the proper volume of the star ($r<R$) in the Jordan frame is
\begin{align}
    M_\text{b} =m_\text{b}N_\text{b}= \int_0^R \ 4\pi r^2 \tilde{\rho}_\text{r} A^3 \left(1-\frac{2\mu}{r} \right)^{-1/2}  \d r .
\end{align}
This is the total energy we would have if we separated all the $N_\text{b}$ baryons in the star into distant places in a thought experiment.

\begin{figure}
    \centering
    \includegraphics[width=\columnwidth]{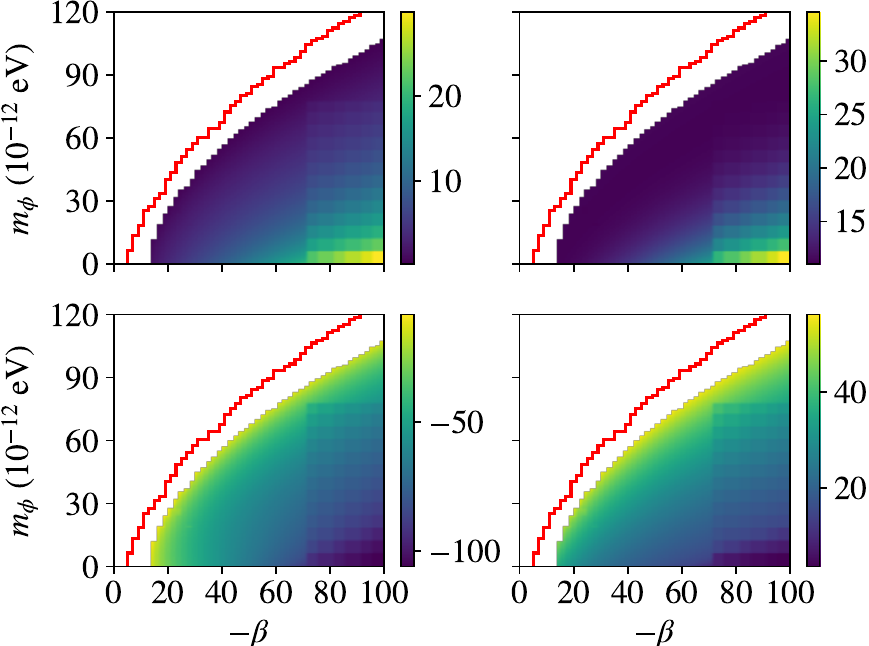}
    \caption{Analog of Fig.~\ref{fig: energy differences HB EOS} for different stellar parameters. Upper row: $\Delta \tilde{R}$, the radius difference between the two stars which have the maximum ADM mass difference for the same baryon mass (left) and the stellar radius at the scalarization branch-off point $\tilde{R}$ (right) in $\textrm{km}$. Lower row: Similar to upper row, but for the central energy density difference $\Delta \tilde{\rho}_\text{c}$ (left) and $\tilde{\rho}_\text{c}$ at the scalarization branch-off point (right) in $10^{16}\,\textrm{kg}/\textrm{m}^3$.}
    \label{fig: radius and rhoc}
\end{figure}

We repeat the above procedure many times to obtain Fig.~\ref{fig: energy differences HB EOS}. In a similar vein, we plot the radius and central density of the stars (in the Jordan frame) in Fig.~\ref{fig: radius and rhoc}. The differences $\Delta \tilde{R}$ and $\Delta \tilde{\rho}_\text{c}$ can be quite prominent. This was an important factor in constraining scalarization via mass-radius observations of NSs in \textcite{Tuna:2022qqr}.

\begin{figure}
    \centering
    \includegraphics[width=\columnwidth]{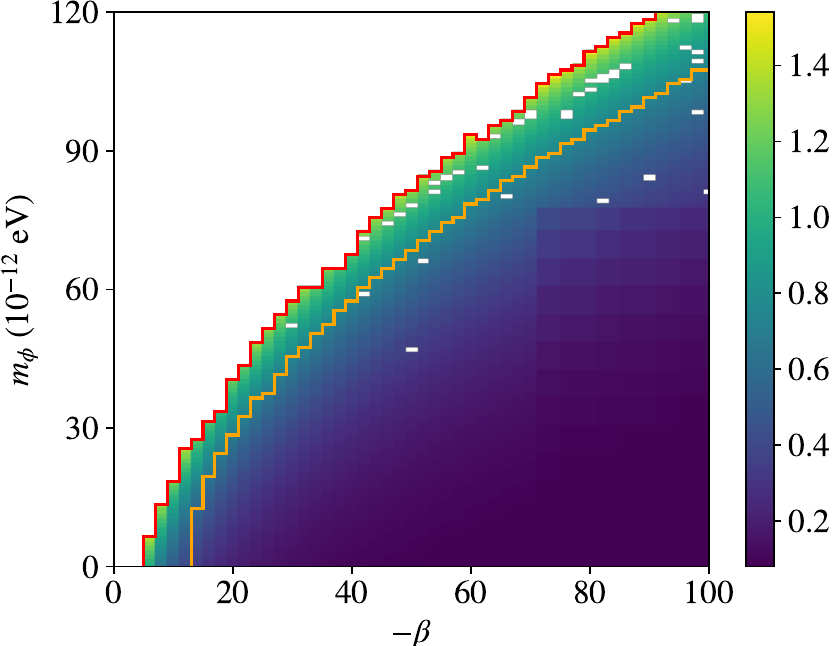}
    \caption{An earlier version of the right subplot of Fig.~\ref{fig: energy differences HB EOS}, where we also give the ADM mass everywhere. Our code failed on the scattered white dots at this stage, which could be filled in by interpolation, but ultimately this was not needed since we had the resources to obtain the solutions eventually. The orange line separates first- and second-order scalarization, the former being characterized by $\Delta M>10^{-5}M_\odot$.}
    \label{fig: white spots}
\end{figure}

Our relaxation scheme sometimes fails to converge to a solution~\cite{Tuna:2022qqr}, and we repeat the procedure with slightly different initial guesses. There might still be failures after many trials, which can be easily filled with interpolation on the $(\beta,\mphi)$ plane. We obtained all the data points eventually, hence did not need interpolation, but this is likely not an optimal use of computational resources. We provide an earlier, incomplete version of Fig.~\ref{fig: energy differences HB EOS}, Fig.~\ref{fig: white spots}, as an example of the possible need for interpolation. 

\noindent {\bf \em Basic aspects of first-order phase transitions:}
Here we summarize how the Landau ansatz of Eq.~\eqref{eq: energy landau ansatz} explains first-order scalarization as seen in Fig.~\ref{fig:first_order}. To start with, for $\chi>1/4$, the only equilibrium point and the global minimum of Eq.~\eqref{eq: energy landau ansatz} is at $Q=0$. The only configuration for a given $M_\text{b}$ is stable and unscalarized, and we are in the region to the left of the line $A$ in Fig.~\ref{fig: binding energy atypical}.

As $M_\text{b}$ increases, $\chi$ drops below $1/4$ and local maxima ($-$) and outer local minima ($+$) appear at
\begin{equation}
\label{eq: local maxima}
Q^2_\pm = \left[-b/(2c)\right]\left(1\pm\sqrt{1-4\ ac/b^2} \ \right).
\end{equation}
However, the \emph{global} minimum is still located at $Q=0$ if $\chi> 3/16$. This is the case where the unscalarized ($Q=0$) solution is the globally stable one, and the stable scalarized solutions are metastable. In Fig.~\ref{fig: binding energy atypical}, this corresponds to the region between vertical lines A and B. The local maxima are also equilibrium solutions, they are the unstable scalarized configurations (dotted line in Fig.~\ref{fig: binding energy atypical}) which have the highest total energy, i.e., the lowest binding energy among the equilibria.

As $M_\text{b}$ increases further, we arrive at $\chi = 3/16$, where the energies at $Q=0$ and the outer minima become equal. This corresponds to line B in Fig.~\ref{fig: binding energy atypical}, and it is the actual point of the first-order phase transition. If $\chi$ becomes less than $3/16$, i.e., to the right of line B, the scalarized solution at $Q_+$ becomes the globally stable one, and the unscalarized stars become metastable. Note that, at the point of transition, the global minimum jumps \emph{discontinuously} from $Q=0$ to one of $Q=\pm Q_+$, which is a hallmark of first-order phase transitions.

Further increasing $M_\text{b}$ brings us to $a=0$ or $\chi=0$, for which the maxima at $Q = \pm Q_-$ join the local minimum at $Q=0$. This corresponds to the line C in Fig.~\ref{fig: binding energy atypical}. For negative values of $a$, hence of $\chi$, we have a local maximum at $Q=0$, and minima at $Q = \pm Q_+$. This corresponds to having one unstable GR solution and one stable scalarized solution, which is the region to the right of line C.

\noindent {\bf \em Extracting the Landau ansatz parameters:}
For fixed $(\beta, m_\phi)$ and EOS, let us expand each parameter in the ansatz~\eqref{eq: energy landau ansatz} as a power series around the critical baryon mass (at the branch-off point), that is,
\begin{align}
\label{eq: coeff expansion}
    a(M_\text{b}) &= \phantom{b_0 + } a_1 (M_\text{b}-M_\text{crit}) + \dots \\
    b(M_\text{b}) &= b_0 + b_1 (M_\text{b}-M_\text{crit}) + \dots
\end{align}
where we used the fact that $a(M_\text{crit})=0$. Consider the scalarized equilibrium solutions of first-order scalarization ($b_0<0$) that are close to the branch-off point, i.e. those stars with $M_\text{b} \approx M_\text{crit}$. These are unstable solutions which correspond to the local maxima in Eq.~\eqref{eq: local maxima}, and their scalar charge $Q=Q_-$ is given by
\begin{equation}
\label{eq: fit Q}
    Q^2 = -\frac{a_1}{b_0} (M_\text{crit} - M_\text{b}) + \mathcal{O}\left[(M_\text{b}-M_\text{crit})^2\right] .
\end{equation}
Hence, the corresponding ADM mass is
\begin{equation}
\label{eq: fit MADM}
    M_\text{ADM} -M_0 = -\frac{1}{2} \frac{a_1^2}{b_0} (M_\text{b}-M_\text{crit})^2 + \dots .
\end{equation}
Combining the last two equations in the leading order, we finally obtain
\begin{equation}
\label{eq: fit b_0}
    2(M_0 -M_\text{ADM}) = b_0\ Q^4 + \dots ,
\end{equation}
which holds well for stars with baryon masses close to $M_\text{crit}$. In summary, we calculate the ADM mass and the scalar charge of a star near the critical baryon mass, as well as $M_0$, which is the ADM mass of an unscalarized ($Q=0$) star with the same baryon mass. If we repeat this for different baryon masses, the $(M_0 - M_\text{ADM})$ versus $Q^4$ curve for small enough $Q$ is simply a line with slope $b_0/2 = b(M_\text{crit})/2$. 

Eq.~\eqref{eq: fit b_0} holds for second order scalarization as well, hence we can perform the same fitting procedure as we change the theory parameters $(\beta,\mphi)$ without worrying about the order of the phase transition. Then, we can obtain $b_0$ as a function of $(\beta,\mphi)$ (and the EOS), whose sign determines the order of scalarization we have on different parts of the STT parameter space. We reported these findings in the main text, see Fig.~\ref{fig:b_vs_beta}.

\begin{figure}
    \centering
    \includegraphics[width=\columnwidth]{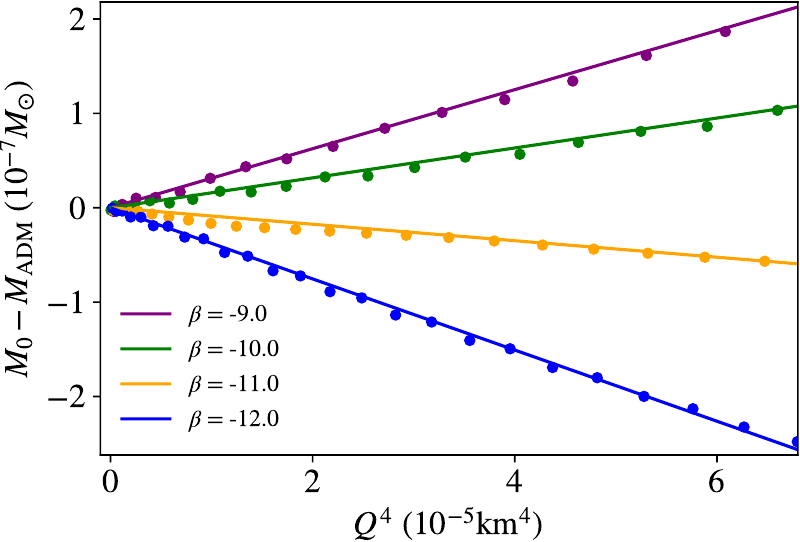}
    \caption{Sample linear fits that yield $b_0$ via Eq.~\eqref{eq: fit b_0} for various values of $\beta$ ($\mphi=0$). The transition from second-order scalarization to first-order occurs around $\beta=-10.6$, where the slope of the fitting line, hence $b_0$, vanishes. See Fig.~\ref{fig:b_vs_beta} for the complete curve $b_0(\beta)$. The data points have small deviations from the fit lines due to numerical errors, and we exclude the points closest to the origin from the fit due to high relative errors. There are further data points outside the frame.}
    \label{fig:deltaM vs Q4}
\end{figure}

The fitting procedure above requires a numerical setup with particularly high precision. $M_\text{ADM}$ and $M_0$ are very close to each other near the critical baryon mass. Hence, when we take their difference, we suffer from the phenomenon of \emph{loss of precision} (also known as \emph{catastrophic cancellation}): The distinct numerical errors in $M_\text{ADM}$ and $M_0$ do not cancel each other, and the relative error grows tremendously due to $|M_\text{ADM}-M_0|$ being exceedingly small compared to either of the masses. We decreased the finite differencing step sizes and the tolerances of the Newton iteration in our relaxation scheme more than an order of magnitude compared to our standard computations, so that the fit is not overwhelmed by the error. See Fig.~\ref{fig:deltaM vs Q4}.

\end{document}